\documentclass[structabstract]{aa}
\usepackage{graphicx}
\usepackage{txfonts}
\begin{document}
   \title{ He star evolutionary channel to intermediate-mass binary pulsar PSR J1802-2124}

   \author{Wen-Cong Chen
     \inst{1,}
     \inst{2,}
     \inst{4}
     \and
      Xiang-Dong Li
     \inst{3,}
     \inst{4}
     \and
     Ren-Xin Xu
      \inst{1}
      }

   \institute{School of Physics and State Key Laboratory of Nuclear Physics
     and Technology, Peking University, Beijing 100871, China
     \and
      Department of Physics, Shangqiu Normal University,
     Shangqiu 476000, China
     \and
     Department of Astronomy, Nanjing University, Nanjing 210093, China
     \and
     Key Laboratory of Modern Astronomy and Astrophysics (Nanjing University), Ministry of Education, Nanjing 210093, China
     \\
     \email{chenwc@nju.edu.cn}
     }

   \date{}


   \abstract
   {The intermediate-mass binary pulsars (IMBPs) are characterized by
relatively long spin periods (10 - 200 ms) and massive ($\ga 0.4~
M_{\odot}$) white dwarf (WD) companions.  Recently, precise mass
measurements have been performed for the pulsar and the WD in the
IMBP PSR J1802-2124. Some observed properties, such as the low
mass of the pulsar, the high mass of the WD, the moderately long
spin period, and the tight orbit, imply that this system has
undergone a peculiar formation mechanism. }
   {In this work,
we attempt to simulate the detailed evolutionary history of PSR
J1802-2124.}
   {We propose that a binary system consisting of a
neutron star (NS, of mass $1.3~ M_{\odot}$) and an He star (of
mass $1.0~ M_{\odot}$), and with an initial orbital period of 0.5
d, may have been the progenitor of PSR J1802-2124. Once the He
star overflows its Roche lobe, He-rich material is transferred
onto the NS at a relatively high rate of $\sim 10^{-7}-10^{-6
}~M_{\odot}\,\rm yr^{-1}$, which is significantly higher than the
Eddington accretion rate. A large amount of the transferred
material is ejected from the vicinity of the NS by radiation
pressure and results in the birth of a mildly recycled pulsar. }
   {Our simulated results
are consistent with the observed parameters of PSR J1802-2124.
Therefore, we argue that the NS + He star evolutionary channel may
be responsible for the formation of most IMBPs with orbital
periods $\la 3~\rm d$. }
   {}

   \maketitle


\section{Introduction}

In the standard recycling theory, a neutron star (NS) in a
low-mass X-ray binary (LMXB) accretes mass from its companion
star, and the transferred angular momentum causes it to be spun up
to a short spin period (Alpar et al. 1982; Bhattacharya \& van den
Heuvel 1991). The endpoint of the evolution is a binary consisting
of a millisecond radio pulsar with spin period $\la 10 ~\rm ms$,
and an He white dwarf (WD). This population is called low-mass
binary pulsars (LMBPs) (Tauris \& van den Heuvel 2006). In
comparison with LMBPs, another population is that of the
intermediate-mass binary pulsars (IMBPs), which contain a pulsar
with a spin period of tens of milliseconds and a massive CO or
ONeMg WD with mass of $\ga 0.4~M_{\odot}$ (Camilo et al. 1996;
Camilo et al. 2001).

It is noticeable that the spin periods, magnetic field strengths,
and orbital eccentricities  of IMBPs are considerably greater than
those of LMBPs (Li 2002), implying that these pulsars are mildly
recycled. So far, the formation mechanisms of IMBPs have not been
fully understood. Van den Heuvel (1994) suggests that these
objects originate in low/intermediate-mass X-ray binaries
(L/IMXBs) with a donor star on the asymptotic giant branch. When
the donor star overflows its Roche-lobe, the unstable mass
transfer onto the NS will lead to the formation of a common
envelope (CE). During the spiral-in process, the NS experiences a
short timescale, super-Eddington accretion and is partially spun
up (\cite{heuv84,stai04}). More recent works indicate that IMBPs
are likely to evolve through a (sub)thermal timescale mass
transfer in IMXBs without experiencing CE evolution (King \&
Ritter 1999; Podsiadlowski \& Rappaport 2000; Kolb et al. 2000;
Tauris et al. 2000; Podsiadlowski et al. 2002; Pfahl et al. 2003.
See also \cite{pyly88}). \cite{li02} proposes that the
thermal-viscous instability of the accretion disks in some IMXBs
may reduce the amount of mass transferred onto the NS and result
in the birth of a mildly recycled pulsar.

At present, 16 known IMBPs exist, and only two of them have
relatively precise measurements of the pulsar masses. For PSR
J0621+1002 the measured mass is $M_{\rm
NS}=1.70^{+0.10}_{-0.16}~M_{\odot}$ (\cite{nice08}).  The other
well-studied IMBP, PSR J1802-2124, was discovered by the Parkes
Multibeam Pulsar Survey (\cite{faul04}). It has a spin period of
12.6 ms and is in a 16.8 hr circular orbit with a massive WD
companion. By measuring the general relativistic Shapiro delay,
\cite{ferd10} estimate the NS mass to be $1.24(\pm 0.11)~
M_{\odot}$ and WD mass to be $0.78(\pm 0.04)~ M_{\odot}$ (68\%
confidence).

Tauris et al. (2000) have calculated the nonconservative evolution
of IMXBs and find that IMBPs with a wide orbit ($P_{\rm orb}\ga
3~\rm d$) can be formed through a short-lived highly
super-Eddington mass transfer phase. However, this scenario may
not work for IMBPs with a short orbital period, such as PSR
J1802-2124, PSRs B0655+64 (\cite{jone88,lori95}), J1232-6501,
J1435-6100 (\cite{cami01,man01}), and J1756-5322 (\cite{edwa01}).
The CO WD may be formed  when the donor star evolves to be a He
star (van den Heuvel 1994), which means the NSs have undergone a
short timescale ($\sim 10^{3}$ yr) accretion from the donor star
in the CE evolution phase (Ferdman et al. 2010). However, the
detailed evolutionary histories of IMBPs with a short orbital
period like PSR J1802-2124 have not been explored, so will be
investigated in this work. Employing the observed parameters of
PSR J1802-2124, we simulate its evolutionary history by using a
detailed stellar evolution code. We argue that a binary system
consisting of an NS and an He star may be its progenitor. In
section 2, we describe the formation processes of NS + He star
systems and obtain the initial parameter space of NS + He star
systems. In section 3 we present a detailed evolutionary path of
PSR J1802-2124. We give a brief summary and discussion in section
4.
\section{Formation of NS + He star systems}

The primordial binaries that formed NS + He star systems consist
of a massive primary and an intermediate-mass secondary that
produce the NS and the WD, respectively. We used an evolutionary
population synthesis approach based on the rapid binary star
evolution (BSE) code developed by Hurley et al. (2000, 2002) to
derive the initial parameter space of NS + He star systems. For
the NS formation, there are two main evolutionary channels: iron
core-collapse (ICC) supernovae (SNe) and electron-capture (EC)
SNe. The former is believed to form a higher mass ($\sim
1.35~M_{\odot}$) population, while the latter produces a lower
mass ($\sim 1.25~M_{\odot}$) population (\cite{pods05};
\cite{schw10}). In calculation, we distinguish ICC SNe of massive
stars from EC SNe or accretion-induced collapse (AIC) of ONeMg
WDs. Some input parameters including the star formation rate, the
initial mass functions, the CE ejection efficiency parameter, the
kick velocity dispersion, the distributions of initial orbital
separations, masses, and the eccentricities are described as
follows\footnote{For other parameters, we refer to Table 3 of
\cite{hurl02} if not mentioned.}.
 \begin{enumerate}
\item We adopt a constant star formation rate $S=7.6085~\rm
yr^{-1}$, which implies that one binary with $M_{1}\geq 0.8~
M_{\odot}$ is born in the Galaxy per year.

\item The primary star mass distribution function is taken to be
$\Phi({\rm ln}M_{1})=M_{1}\xi(M_{1})$, in which the initial mass
function $\xi(M_{1})$ is given by \cite{krou93}.

\item The secondary star mass distribution function is taken to be
$\Phi({\rm ln}M_{2})=M_{2}/M_{1}=q$, which corresponds to a
uniform distribution of the mass ratio $q$ between 0 and
1\footnote{Recently, \cite{pins06} proposed that two components in
50\% of detached binaries have similar masses, i.e. $q>0.87$.}.

\item We take a uniform distribution of ${\rm ln}a$ for the binary
separation $a$, namely $\Phi({\rm ln}a)=0.12328$ (Hurley et al.
2002).

\item The CE ejection efficiency parameter $\alpha_{\rm CE}$,
which describes the fraction of the released orbital energy used
to eject the envelope during the CE evolution, i. e. $\alpha_{\rm
CE}=E_{\rm bind}/(E_{\rm orb,f}-E_{\rm orb,i})$ (Hurley et al.
2002). Here $E_{\rm bind}$ is the total binding energy of the
primary's envelope, $E_{\rm orb,f}$, and $E_{\rm orb,i}$ are the
final and initial orbital energies of the secondary, respectively.
In this work, we adopt $\alpha_{\rm CE}=3$ in our standard model
and $\alpha_{\rm CE}=1$ for comparison.

\item The kick velocity dispersion received by the natal NS. For
NSs formed by ICC SNe, we adopt relatively stronger kick
velocities with a dispersion $\sigma_{\rm ICC}=190~\rm km\,s^{-1}$
(\cite{hans97})\footnote{\cite{hobb05} find that the observed
pulsars could be described by a Maxwellian distribution with a
dispersion $\sigma=265~\rm km\,s^{-1}$. However, the population
synthesis by \cite{belc10} suggests that, only when the kick
velocity dispersion of natal NS formed in close interacting
binaries is $\sim170~\rm km\,s^{-1}$, can the simulated results
fit the observed intrinsic ratio between disrupted recycled
pulsars and double NSs.}. However, for the EC SNe or AIC channel,
the explosive energies of electron-capture SN explosion are
significantly lower than those of ICC SNe (\cite{dess06};
\cite{kita06}; \cite{jian07}). Therefore, it is usually believed
that the natal NSs formed by both channels may receive a smaller
kick (\cite{pods04}; \cite{sche04}; \cite{poel08}; for a review,
see \cite{heuv10}). Therefore, we take a moderately weak velocity
dispersions with $\sigma_{\rm EC,AIC}=20$ km\,s$^{-1}$
\footnote{For the natal NS via EC SN, \cite{dewi06} proposes that
the kick is a Maxwellian distribution with a dispersion of 30 $\rm
km \,s^{-1}$.}.

\item All stars are assumed to be members of binaries  in a
circular orbit, and they have metallicity of $Z=0.02$. In our
simulations, the primary mass $M_{1}$, the secondary mass $M_{2}$,
and the binary separation $a$ range in $0.8-80M_{\odot}$,
$0.1-80M_{\odot}$, and $3-10000R_{\odot}$, respectively.
 \end{enumerate}

First, we show an example of the formation process of an NS + He
star system. Using the BSE code we calculated the evolution of a
primordial binary consisting of a primary (star 1) of mass
$M_{1}=8.5~M_{\odot}$ and a secondary (star 2) of mass
$M_{2}=4.4~M_{\odot}$ ($Z=0.02$) in an $a=100.7~R_{\odot}$ orbit.
At $t=33.11~\rm Myr$ the primary evolves to fill its Roche lobe,
and transfers its material to star 2. Because of the relatively
high-mass ratio, the mass transfer is dynamically unstable,
resulting in the formation of a CE. After the envelope is ejected,
the primary becomes an He star ($M_{1}=1.738~M_{\odot}$), the
donor star is still on the main sequence
($M_{2}=6.837~M_{\odot}$), but the orbit shrinks to
$a=33.19~R_{\odot}$. At $t=40.03~\rm Myr$, the He star overflows
its Roche lobe once more, and the product of the subsequent
evolution is a binary consisting of an ONeMg WD
($M_{1}=1.035~M_{\odot}$) and a main sequence star
($M_{2}=7.456~M_{\odot}$), in an orbit of $a=70.915~R_{\odot}$. At
$t=71.77~\rm Myr$, Star 2 starts to overflow its Roche lobe,
initiating the second CE evolution. After the CE is ejected, a
close binary system consisting of an ONeMg WD
($M_{1}=1.035~M_{\odot}$) and an He star ($M_{2}=1.459~M_{\odot}$)
is formed in a tight orbit with $a=0.725~R_{\odot}$ (see also
\cite{belc04}). When the He star (the descendant of Star 2) fills
its Roche lobe and transfers the He-rich material onto the ONeMg
WD, the WD may grow to the Chandrasekhar limit and collapse to be
an NS because of the electron capture process (\cite{nomo84})
\footnote{Here, we present an evolutionary example of an NS + He
star system. However, the NS may also be formed directly by an
electron-capture supernova of the He-star that originated from a
close binary by Case B mass transfer, and its progenitor star
should have a mass in the range 8 to 12 $M_{\odot}$ (Podsiadlowski
et al. 2004).}. Therefore, through two CE phases a compact binary
including an NS and a low-mass He star ($M_{2}=1.049~M_{\odot}$)
can be produced. Meanwhile, with the sudden increase in the
orbital separation caused by the received kick during the AIC, the
mass transfer ceases.

To investigate the initial parameter space of NS + He star
systems, we calculated the evolution of $1\times 10^{6}$ binaries
to an age of 12 Gyr. Figure 1 shows the distribution of natal NS +
He star systems in the $M_{\rm He}-P_{\rm orb}$ diagram when
$\sigma_{\rm EC, AIC}=20~\rm km\,s^{-1}$. It is clear that most
systems have initial orbital periods of $P_{\rm orb,i}\sim0.01 - 1
~\rm d$ and initial He star masses $M_{\rm He,i}\sim 0.5 -
3.0~M_\odot$. For the same kick, a lower $\alpha_{\rm CE}$ results
in fewer NS + He star systems. This difference develops out of the
influence of $\alpha_{\rm CE}$ on the evolution of close binaries.
A high $\alpha_{\rm CE}$ can prevent coalescence during the CE
stage, significantly increasing the formation rate of NS + He star
systems (Liu \& Li 2006).

\begin{figure}
\centering
\includegraphics[angle=0,width=9cm]{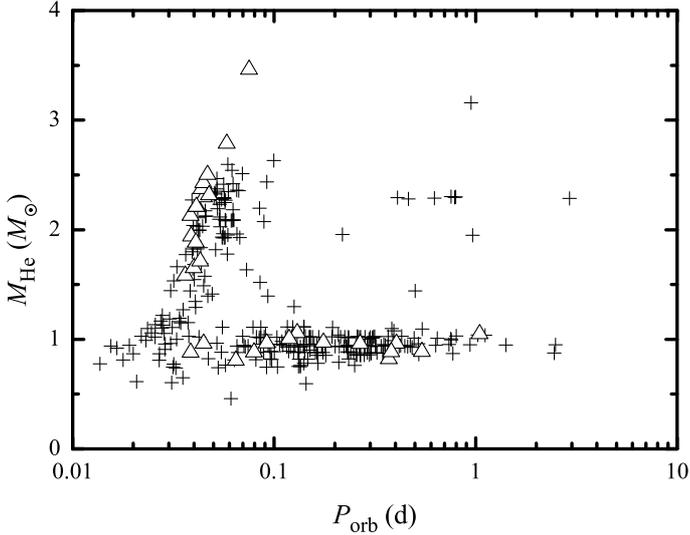}
\caption{Distribution of initial NS + He star systems in the
$M_{\rm He}-P_{\rm orb}$ diagram when  $\sigma_{\rm ICC}=190~\rm
km\,s^{-1}$, and $\sigma_{\rm EC, AIC}=20~\rm km\,s^{-1}$. The
crosses and the open triangles correspond to the CE ejection
efficiency parameters $\alpha_{\rm CE}=3$, and 1, respectively.}
\label{Fig1}
\end{figure}

\section{An example evolutionary path to PSR J1802-2124}

In this section we show an example of an evolutionary path to PSR
J1802-2124, to reproduce its observed parameters. We used the
stellar evolution code developed by Eggleton (1971,1972,1973) to
calculate the evolutionary sequences of an NS + He star system. In
the calculations, we adopted the stellar OPAL opacities for a low
temperature given by \cite{roge92} and \cite{alex94} and take the
ratio of the mixing length to the pressure scale height to be 2.0.
The overshooting parameter of the He star (with a chemical
abundance $Y = 0.98$,  $ Z = 0.02$) is set to 0 (\cite{wang09}).

When the He star overflows its Roche lobe and transfers He-rich
material onto the NS, the maximum accretion rate of the NS should
be limited by the Eddington accretion rate, which is twice that of
hydrogen accretion, i. e. $\dot{M}_{\rm Edd}\simeq 3.0\times
10^{-8}~M_{\odot}{\rm yr}^{-1}$. If the mass transfer rate
$-\dot{M}_{\rm He}$ from the He star is greater than $\dot{M}_{\rm
Edd}$, the radiation pressure from the NS will cause the excess
mass to be lost from the system at a rate $\dot{M}=\dot{M}_{\rm
He}+\dot{M}_{\rm Edd}$. In our calculations, we consider three
types of orbital angular momentum loss from the binary system: (1)
gravitational-wave radiation, (2) magnetic braking: we adopt the
induced magnetic braking prescription given by \cite{sill00}, (3)
isotropic winds: the transferred matter in excess of the Eddington
accretion rate is assumed to be ejected from the vicinity of the
NS in the form of isotropic winds and to carry away the specific
orbital angular momentum of the NS (\cite{taur99}).

Based on the results of Figure 1, we calculated the evolution of
an NS + He star system with initial masses $M_{\rm
NS,i}=1.3~M_{\odot}$, $M_{\rm He,i}=1.0~M_{\odot}$, and initial
orbital period $P_{\rm orb,i}=0.5~\rm d$. Figures 2 and 3 show the
evolution of the mass transfer rate, the NS mass, the orbital
period, and the He star mass. At $t\simeq 17.71\rm Myr$, the He
star fills its Roche lobe and transfers mass
 at a high rate of $\sim 10^{-7}-10^{-6 }~M_{\odot}\,\rm
yr^{-1}$ onto the NS, which is significantly higher than the
Eddington limit. A large amount of the transferred material is
ejected by radiation pressure, so the mass of the NS hardly gains
anything. Because the material is transferred from the less
massive He star to the more massive NS, the orbital period
continuously increases. The mass exchange lasts $\sim 0.25~\rm
Myr$, during which the NS has accreted $\sim 4\%$ of the
transferred mass, or $\sim 0.01~M_{\odot}$. This amount of
material is, however, sufficient to spin the NS up to $\la 16$ ms
and cause the magnetic field to decay (\cite{taam86}; van den
Heuvel 1994). The endpoint of the evolution is an IMBP consisting
a mildly recycled pulsar and a CO WD with a mass of
$0.81~M_{\odot}$ in an orbital period of $0.71~\rm d$. This
evolution provides a plausible path that leads to the formation of
PSR J1802-2124.

\begin{figure}
\centering
\includegraphics[angle=0,width=9cm]{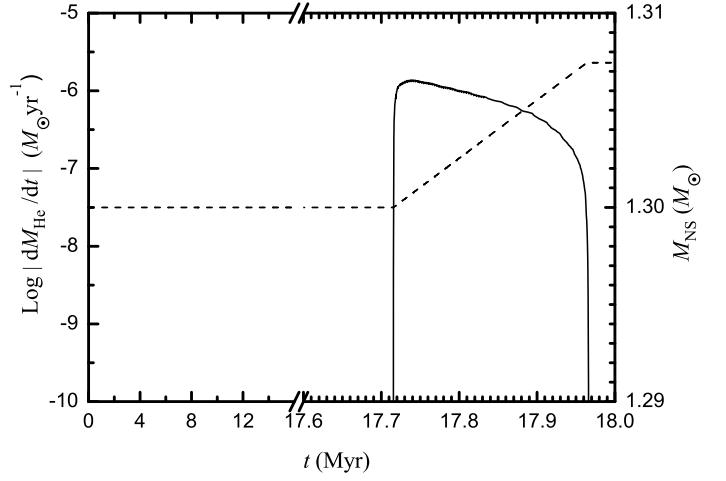}
\caption{Evolution of an NS + He star system with $M_{\rm
NS,i}=1.3~M_{\odot}$, $M_{\rm He,i}=1.0~M_{\odot}$, and $P_{\rm
orb,i}=0.5~\rm d$. The solid and dashed curves denote the
evolutionary tracks of the mass transfer rate and the NS mass,
respectively.} \label{Fig1}
\end{figure}

\begin{figure} \centering
\includegraphics[angle=0,width=9cm]{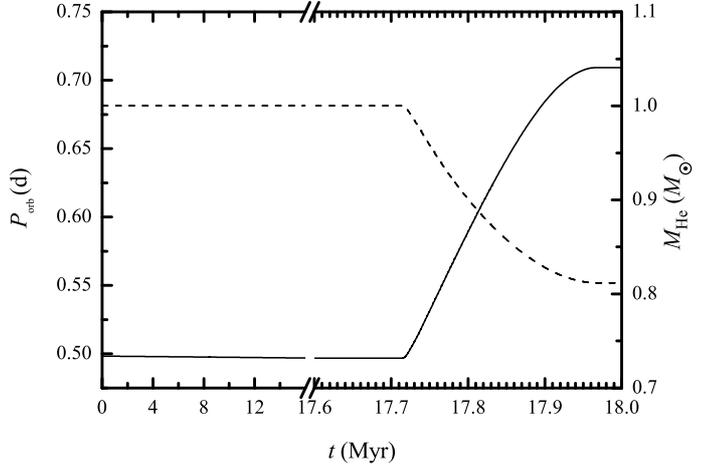}
\caption{Evolution of an NS + He star system with $M_{\rm
NS,i}=1.3~M_{\odot}$, $M_{\rm He,i}=1.0~M_{\odot}$, and $P_{\rm
orb,i}=0.5~\rm d$. The solid and dashed curves denote the
evolutionary tracks of the orbital period and the He star mass,
respectively.} \label{Fig1}
\end{figure}

\section{Discussion and summary}
At present, the formation channel of IMBPs is not very well
understood. The recent timing analysis of PSR J1802-2124 provides
precise measurements of the pulsar and WD masses (Ferdman et al.
2010). The observed properties of this system, such as the low
mass of the pulsar, the high mass of the WD, short orbital period,
and moderately long spin period (in comparison with those of
LMBPs) imply that this binary may have a specific evolutionary
history. One possibility is that IMBPs like PSR J1802-2124 may
have evolved from IMXBs. Because the material is transferred from
the more massive donor star to the less massive NS in IMXBs, the
mass transfer occurs on a short thermal timescale. Therefore,
\cite{pods02} suggest that IMXBs spend most of their X-ray active
lifetime as LMXBs. In this phase some of the binaries would have
undergone unstable accretion owing to an accretion disk
instability during the LMXB phase, and thus the accretion
efficiency was significantly decreased (Li 2002).

In this work, we attempted to explore whether the evolutionary
channel of NS + He star system can reproduce the observed
parameters of PSR J1802-2124. We first investigated the initial
parameter space of NS + He star systems using a population
synthesis approach. Our simulated results show that most NS + He
star systems have initial orbital periods $P_{\rm orb,i}\sim0.01 -
1 ~\rm d$ and initial He star masses $M_{\rm He,i}\sim 0.5 -
3.0~M_\odot$. We then performed numerical calculations for the
evolution of an NS + He star system with $M_{\rm
NS,i}=1.3~M_{\odot}$, $M_{\rm He,i}=1.0~M_{\odot}$, and $P_{\rm
orb,i}=0.5~\rm d$. Detailed evolutionary calculations shows that
the NS + He star evolutionary channel can successfully reproduce
the observed parameters of PSR J1802-2124. The NS + He star
systems have a compact orbit, therefore we propose that the NS +
He star evolutionary channel may be responsible for the formation
of most IMBPs with orbital periods $\la 3~\rm d$.

Obviously, two main uncertainties exist in our work. The first one
is the CE evolution. As seen in Section 2, the initial parameter
space of the NS + He star systems strongly depends on the CE
ejection efficiency parameters $\alpha_{\rm CE}$, which should
vary with stellar mass and evolution, but is still poorly known.
Additionally, some works argue that, an NS may experience
hypercritical accretion in the CE phase and collapse to be a black
hole (see \cite{chev93,brow95,brow01}). However, it has been found
that $\sim 60\%$ NS with a mass greater than $2.0~M_{\odot}$ can
survive under hypercritical accretion (\cite{belc02}). The second
one is the EC SNe and AIC process. When the ONeMg core of an
asymptotic giant branch star (\cite{sies07}; \cite{poel08}) or an
He star (\cite{nomo87}) grows to a critical mass, EC SN would be
triggered. However, the AIC process originates from an ONeMg WD
accreting from the donor star or from the merger of two CO WDs
(\cite{nomo85}; \cite{nomo91}). Compared with the AIC channel, EC
SN can also produce a detached NS + He star system. To obtain a
long enough spin-down timescale for the natal NS, our evolutionary
scenario favors detached binary systems. Therefore, it seems that
also NS formed by EC SN are reasonable progenitors of mildly
recycled pulsars. Podsiadlowski et al. (2004) suggest that the
minimum mass of the NS progenitor may be $10-12~M_{\odot}$ for
single stars, while this value can be $6-8~M_{\odot}$ in binaries.
The progenitor masses of EC SNe are related to the amount of
convective overshooting, and to the metallicity of the star
(\cite{pods04}). A full population synthesis for the birthrates of
NS + He star systems formed by EC SNe and AIC channels is beyond
the scope of this paper, and will be addressed in future work.

\begin{acknowledgements}
We thank the anonymous referee for his/her valuable comments that
helped us to improve the paper. This work was supported by the
National Science Foundation of China (under grant numbers
10873008, 10873011, and 10973002), the National Basic Research
Program of China (973 Program 2009CB824800), China Postdoctoral
Science Foundation funded project, Program for Science \&
Technology Innovation Talents in Universities of Henan Province,
and Innovation Scientists and Technicians Troop Construction
Projects of Henan Province, China.
\end{acknowledgements}


\begin{thebibliography}{}
\bibitem[Alexander \& Ferguson (1994)]{alex94} Alexander, D. R., \& Ferguson, J. W. 1994, ApJ, 437, 879
\bibitem[Alpar et al. 1982]{alpa82} Alpar, M. A., Cheng, A. F., Ruderman, M. A. \&  Shaham, J. 1982, Nature, 300, 728
\bibitem[Belczynski et al. 2002]{belc02} Belczynski, K., Kalogera, V., \& Bulik, T. 2002, ApJ, 572, 407
\bibitem[Belczynski \& Taam 2004]{belc04} Belczynski, K., \& Taam, R. E. 2004, ApJ, 603, 690
\bibitem[Belczynski et al. (2010)]{belc10} Belczynski, K., Lorimer, D. R., Ridley, J. P., \& Curran, S. J. 2010, MNRAS, 407, 1245
\bibitem[Bhattacharya \& van den Heuvel 1991]{bhat91} Bhattacharya, D., \& van den Heuvel, E. P. J. 1991, Phys. Rep., 203, 1
\bibitem[Brown 1995]{brow95} Brown, G. E. 1995, ApJ, 440, 270
\bibitem[Brown et al. 2001]{brow01} Brown, G. E., Lee, C.-H., Portegies Zwart, S. F., \& Bethe, H. A. 2001, ApJ, 547, 345
\bibitem[Camilo et al. 1996]{cami96} Camilo, F., Nice, D. J., Shrauner, J. A., \& Taylor, J. H. 1996, ApJ, 469, 819
\bibitem[Camilo et al. 2001]{cami01} Camilo, F., et al. 2001, ApJ, 548, L187
\bibitem[Chevalier 1993]{chev93} Chevalier, R. A. 1993, ApJ, 411, L33
\bibitem[Dessart et al. 2006]{dess06} Dessart, L., Burrows, A., Livine, E., \& Ott, C. D. 2006, ApJ, 644, 1043
\bibitem[Dewi et al. (2006)]{dewi06} Dewi, J. D. M., Podsiadlowski, P., \& Sena, A. 2006, MNRAS, 368,1742
\bibitem[Edwards \& Bailes 2001]{edwa01} Edwards, R. T., \& Bailes, M. 2001, ApJ, 553, 801
\bibitem[Eggleton (1971)]{egg71} Eggleton, P. P. 1971, MNRAS, 151, 351
\bibitem[Eggleton (1972)]{egg72} Eggleton, P. P. 1972, MNRAS, 156, 361
\bibitem[Eggleton (1973)]{egg73} Eggleton, P. P. 1973, MNRAS, 163, 279
\bibitem[Faulkner et al. 2004]{faul04} Faulkner, A. J., et al. 2004, MNRAS, 355, 147
\bibitem[Ferdman et al. (2010)]{ferd10} Ferdman, R. D., et al. 2010, ApJ, 711, 764
\bibitem[Hansen \& Phinney 1997]{hans97} Hansen, B. M. S., \& Phinney, E. S. 1997, MNRAS, 291, 569
\bibitem[Hobbs et al.  (2005)]{hobb05} Hobbs, G., Lorimer, D. R., Lyne, A. G., \& Kramer, M., 2005, MNRAS, 360, 974
\bibitem[Hurley et al. (2000)]{hurl00} Hurley, J. R., Pols, O. R., \& Tout, C. A. 2000, MNRAS, 315, 543
\bibitem[Hurley et al. (2002)]{hurl02} Hurley, J. R., Tout, C. A., \& Pols, O. R. 2002, MNRAS, 329, 897
\bibitem[Hurley et al. (2010)]{hurl10} Hurley, J. R., Tout, C. A., Wickramasinghe, D. T., Ferrario, L., Kiel, P. D., 2010, MNRAS, 402, 1437
\bibitem[Jiang et al. 2007]{jian07} Jiang, B., Chen, Y., \& Wang, Q. D. 2007, ApJ, 670, 1142
\bibitem[Jones \& Lyne 1988]{jone88} Jones, A. W., \& Lyne, A. G. 1988, MNRAS, 232, 473
\bibitem[Kitaura et al. 2006]{kita06} Kitaura, F. S., Janka, H.-Th., \& Hillebrandt, W. 2006, A\&A, 450, 345
\bibitem[King \& Ritter (1999)]{king99}King, A. R. \& Ritter, H. 1999, MNRAS, 309, 253
\bibitem[Kolb et al. (2000)]{kolb00} Kolb, U., Davies, M. B., King, A. R., \& Ritter, H. 2000, MNRAS,
317, 438
\bibitem[Kroupa et al. (1993)]{krou93} Kroupa, P., Tout, C. A., \& Gilmore, G. 1993, MNRAS, 262, 545
\bibitem[Li (2002)]{li02} Li, X. -D. 2002, ApJ, 564, 930
\bibitem[Liu \& Li (2006)]{liu06} Liu, X. -W., \& Li, X. -D. 2006, A\&A, 449, 135
\bibitem[Lorimer et al. 1995]{lori95} Lorimer, D. R., Yates, J. A., Lyne, A. G., \& Gould, D. M. 1995, MNRAS, 273, 411
\bibitem[Manchester 2001]{man01} Manchester, R. N., et al. 2001, MNRAS, 328, 17
\bibitem[Nice et al. 2008]{nice08} Nice, D. J., Stairs, I. H., \& Kasian, L. E. 2008, in AIP Conf.
Ser. 983, 40 Years of Pulsars: Millisecond Pulsars, Magnetars and
More, ed. C. Bassa, Z.Wang, A. Cumming, \& V. M. Kaspi (Melville,
NY: AIP), 453
\bibitem[Nomoto 1984]{nomo84} Nomoto, K. 1984, ApJ, 277, 791
\bibitem[Nomoto 1987]{nomo87} Nomoto, K. 1987, ApJ, 322, 206
\bibitem[Nomoto \& Iben 1985]{nomo85}  Nomoto, K., \& Iben, I., Jr. 1985, ApJ, 297, 531
\bibitem[Nomoto \& Kondo 1991]{nomo91}  Nomoto, K., \& Kondo, Y. 1991, ApJ, 367, L19
\bibitem[Pfahl, Rappaport, \& Podsiadlowski (2003)]{pfahl03}Pfahl, E., Rappaport, S., \& Podsiadlowski, P. 2003, ApJ, 597, 1036
\bibitem[Pinsonneault \& Stanek (2006)]{pins06} Pinsonneault, M. H., \& Stanek, K. Z. 2006, ApJ, 639, L67
\bibitem[Podsiadlowski \& Rappaport (2000)]{pods00}Podsiadlowski, Ph. \& Rappaport, S. 2000, ApJ, 529, 946
\bibitem[Podsiadlowski et al. (2002)]{pods02} Podsiadlowski, Ph., Rappaport, S., \& Pfahl, E. 2002, ApJ, 565, 1107
\bibitem[Podsiadlowski et al. 2004]{pods04} Podsiadlowski, Ph., Langer, N., Poelarends, A. J. T., Rappaport, S., Heger, A., \& Pfahl, E. 2004, ApJ, 612, 1044
\bibitem[Podsiadlowski et al. 2005]{pods05} Podsiadlowski, Ph., Dewi, J. D. M., Lesaffre, P., Miller, J. C., Newton, W. G., \& Stone, J. R. 2005, MNRAS, 361, 1243
\bibitem[Poelarends et al. 2008]{poel08} Poelarends, A. J. T., Herwig, F., Langer, N., \& Heger, A. 2008,
ApJ, 675, 614
\bibitem[Pylyser \& Savonije 1988]{pyly88} Pylyser, E. H. P., \& Savonije, G. J. 1988, A\&A, 208, 52
\bibitem[Rogers \& Iglesias (1992)]{roge92} Rogers, F. J., \& Iglesias, C. A. 1992, ApJS, 79, 507
\bibitem[Scheck et al.  2004]{sche04} Scheck, L., Plewa, T., Janka, H.-Th., Kifonidis, K., \& Muller, E.
2004, Phys. Rev. Lett., 92, 1103
\bibitem[Schwab et al.  2010]{schw10} Schwab, J., Podsiadlowski, Ph., \& Rappaport, S. 2010, ApJ, 719, 722
\bibitem[Sills et al. (2000)]{sill00} Sills, A., Pinsonneault, M. H., \& Terndrup, D. M. 2000, ApJ, 534, 335
\bibitem[Siess 2007]{sies07} Siess, L. 2007, A\&A, 476, 893
\bibitem[Stairs 2004]{stai04} Stairs, I. H., 2004, Science, 304, 547
\bibitem[Taam \& van den Heuvel 1986]{taam86} Taam, R. E., \& van den Heuvel, E. P. J. 1986, ApJ, 305, 235
\bibitem[Tauris \& Savonije 1999]{taur99} Tauris, T. M., Savonije, G. J. 1999, A\&A, 350, 928
\bibitem[Tauris et al. (2000)]{taur00} Tauris, T., van den Heuvel, E. P. J., \& Savonije, G. J. 2000, ApJ, 530, L93
\bibitem[Tauris \& van den Heuvel (2006)]{taur06} Tauris, T. M., \& van den Heuvel, E. P. J. 2006, in Formation and
Evolution of Compact Stellar X-ray Sources, ed. W. H. G. Lewin \&
M. van der Klis (Cambridge: Cambridge Univ. Press), 623
\bibitem[van den Heuvel \& Taam 1984]{heuv84} van den Heuvel, E. P. J., \& Taam, R. E. 1984, Nature, 309, 235
\bibitem[van den Heuvel (1994)]{heuv94} van den Heuvel, E. P. J. 1994, A\&A, 291, L39
\bibitem[van den Heuvel 2010]{heuv10} van den Heuvel, E. P. J. 2010, New Astronomy Reviews, 54, 140
\bibitem[Wang et al. 2009]{wang09} Wang, B., Meng, X., Chen, X., \& Han, Z. 2009, MNRAS, 395, 847


\end{thebibliography}
\end{document}